\documentstyle[prb,twocolumn,aps,psfig,amssymb,floats]{revtex}

\begin{document}
\tolerance 50000
%%%%\preprint{
%%%%\begin{minipage}[t]{1.8in}
%%%%\hfill LPQTH-94/?
%%%%\end{minipage}
%%%%}

\draft

\twocolumn[\hsize\textwidth\columnwidth\hsize\csname @twocolumnfalse\endcsname

\title{Spin dynamics of the spin-Peierls compound $CuGeO_3$ under magnetic 
field}
\author{D. Poilblanc$^{1}$, J. Riera$^{2}$, 
C. A. Hayward$^{1}$, C. Berthier$^{3,4}$ \& M. Horvati\'{c}$^{3}$
}
\address{
$^{1}$Laboratoire de Physique Quantique, Universit\'e Paul Sabatier, 
31062 Toulouse, France \\
$^{2}$Instituto de Fisica Rosario, Universidad National de Rosario,
Bv. 27 de Febrero 210 bis - 2000 Rosario, Argentina \\
$^3$Grenoble High Magnetic Field Laboratory,
Max-Planck-Institut f\"{u}r Festk\"{o}rperforschung and \\
Centre National de la Recherche Scientifique,
BP 166, 38042 Grenoble Cedex 9, France \\
$^4$Laboratoire de Spectrom\'{e}trie Physique,
Unit\'{e} Mixte de Recherche No. 5588 du
Centre National de la Recherche Scientifique,
Universit\'{e} Joseph Fourier Grenoble I, BP 87,
38402 St.-Martin d'H\`{e}res Cedex, France
}

\date{December 96}
\maketitle

\begin{abstract}
\begin{center}
\parbox{14cm}{
The magnetic field--driven transition in the spin-Peierls 
system CuGeO$_3$ associated with the closing of the spin gap is
investigated numerically. The field dependence of the 
spin dynamical structure factor (seen by inelastic neutron scattering) 
and of the momentum dependent static susceptibility are calculated.
In the dimerized phase ($H<H_c$), we suggest that the strong field dependence 
of the transverse susceptibility could be experimentally seen from the
low temperature spin-echo relaxation rate $1/T_{2G}$ or the second moment 
of the NMR spectrum.
Above $H_c$ low energy spin excitations appear at incommensurate wave vectors
where the longitudinal susceptibility $\chi_{zz}(q)$ peaks. 
}
\end{center}
\end{abstract}

\pacs{
\hspace{1.9cm}
PACS numbers: 74.72.-h, 71.27.+a, 71.55.-i}
\vskip2pc]

Low dimensional quantum magnets have drawn great attention in recent years.
It is believed that the inorganic compound CuGeO$_3$ is a good realisation 
of quasi-one dimensional (1D) weakly coupled spin--$1/2$ antiferromagnetic
chains. At low temperature, a Spin-Peierls (SP) transition was experimentally 
established from a rapid drop of the spin susceptibility \cite{Suscep}.
Simultaneously with the opening of a spin gap, the 
lattice undergoes a structural change characterised by a small 
distortion along the chains direction (X--axis) as seen by x-rays diffraction
studies \cite{x_rays}. 

The Peierls instability in S=1/2 Heisenberg chains has, on the theoretical 
side, been a long standing problem \cite{theory_old} first connected to 
experiments on organic quasi-1D systems such as 
(TTF)CuS$_4$C$_4$(CF$_3$)$_4$ 
(Ref. \onlinecite{exp_old}) where TTF stands for tetrathiafulvalene.
The new CuGeO$_3$ compound is the fist {\it inorganic} quasi-1D SP 
system with a critical temperature $T_{SP}\sim 14K$.

To describe a single $Cu^{2+}$ spin-1/2 chain of CuGeO$_3$ 
we start with the following effective hamiltonian \cite{comment_interchain},
\begin{eqnarray}
   {\cal H}=\sum_{i} J(i)\
   {\bf S}_{i} \cdot {\bf S}_{i+1} 
   +J^\prime\;\sum_{i} 
   {\bf S}_{i} \cdot {\bf S}_{i+2} \;    ,
\label{hamiltonian} 
\end{eqnarray}
\noindent
where $i$ is a site index ($i$=1,...,$L$) and ${\bf S}_{i}$ 
is an electron spin operator. A small dimerisation of the nearest neighbor (NN)
exchange coupling is due to the lattice 
distortion and mimics the effect of the phonons, 
$J(i)=J(1+\delta (-1)^i)$. 

Simple quantum chemistry arguments\cite{Khomskii} suggest that, due to side
groups and in contrast to the usual case, the (almost)
$90^o$ Cu-O-Cu superexchange 
path is antiferromagnetic (i.e. $J>0$) with $J\sim 140K$ ($11.6$ meV). 
In addition, the second nearest neighbor exchange coupling $J^\prime$ is also 
antiferromagnetic and hence leads to frustration. 
A sizeable ratio $\alpha=J^\prime/J$ is suggested by the
magnetic susceptibility measurements \cite{Suscep} above $T_{SP}$ ($\delta=0$)
which can be well fitted by parameters like $\alpha=0.24$ and 
$J=150K$ (Ref. \onlinecite{Castilla}) or $\alpha=0.36$ and 
$J=160K$ (Ref. \onlinecite{Riera_suscep}). 

In the dimer (D) or Spin-Peierls phase, for temperatures 
below $T_{SP}\sim 14K$, the chains 
become dimerized with an alternation of somewhat shorter and longer Cu-Cu 
bonds along the X--axis. Physically, this corresponds to the formation
of singlet dimers on the stronger bonds \cite{Majumdar}. 
In the simple ($\delta=0$) $J$--$J^\prime$ Heisenberg chain the 
transition towards a gapped disordered ground state (GS) occurs at
$\alpha_c\sim 0.241$ \cite{critical_value}. The parameters proposed by
Castilla {\it et al.} \cite{Castilla} hence correspond almost exactly to the 
critical point and, at low temperature, an extra coupling to the lattice
must be advocated to give the correct zero temperature finite spin gap
seen in neutron inelastic scattering \cite{Regnault}.
This is taken care of by an {\it ad hoc} dimerisation 
$\delta=0.03$ \cite{Castilla}. In addition to the value of the spin gap
such parameters seem also to reproduce satisfactorilly the whole dynamical 
spin structure factor \cite{Haas}. 
However, it should be noted that a similar agreement on the gap value 
can also be achieved
using the set of parameters of Ref. \protect\onlinecite{Riera_suscep} provided 
a smaller $\delta=0.014$ is used \cite{Riera_field}. 
In any case, it is clear that both magnetic frustration and 
spin-phonon interactions conspire to the formation of the SP phase \cite{DMRG}.
In the following, we shall use both sets of parameters. 

The small value of the spin gap ($\Delta_S\sim 2.1\, \text{meV}$) 
makes the magnetic 
field properties of CuGeO$_3$ particularly interesting. The Zeeman energy
$-g\mu_B S_z$ leads, for $T<T_{SP}$ and for increasing magnetic field 
($H_c\sim 12.5\, T$ at low temperature), to a phase 
transition to an incommensurate or solitonic phase \cite{X_ray2,Berthier,IR}.
This transition is directly connected to closing of the spin gap by the
magnetic field \cite{Cross_field}. An accurate theoretical description of the
full spin {\it dynamics} under magnetic field is still missing and
is necessary for the 
interpretation of neutron inelastic scattering experiments under 
magnetic field or the NMR experiments. In the following, we investigate 
the magnetic field behavior of the dynamical spin structure factor and
of the static q-dependent magnetic susceptibility which can be
indirectly studied in NMR experiments. Dynamical correlations are calculated
on finite clusters by standard exact diagonalisation 
methods \cite{review_numerics}.
\vskip -1.0cm 
\begin{figure}[htb]
\begin{center}
\mbox{\psfig{figure=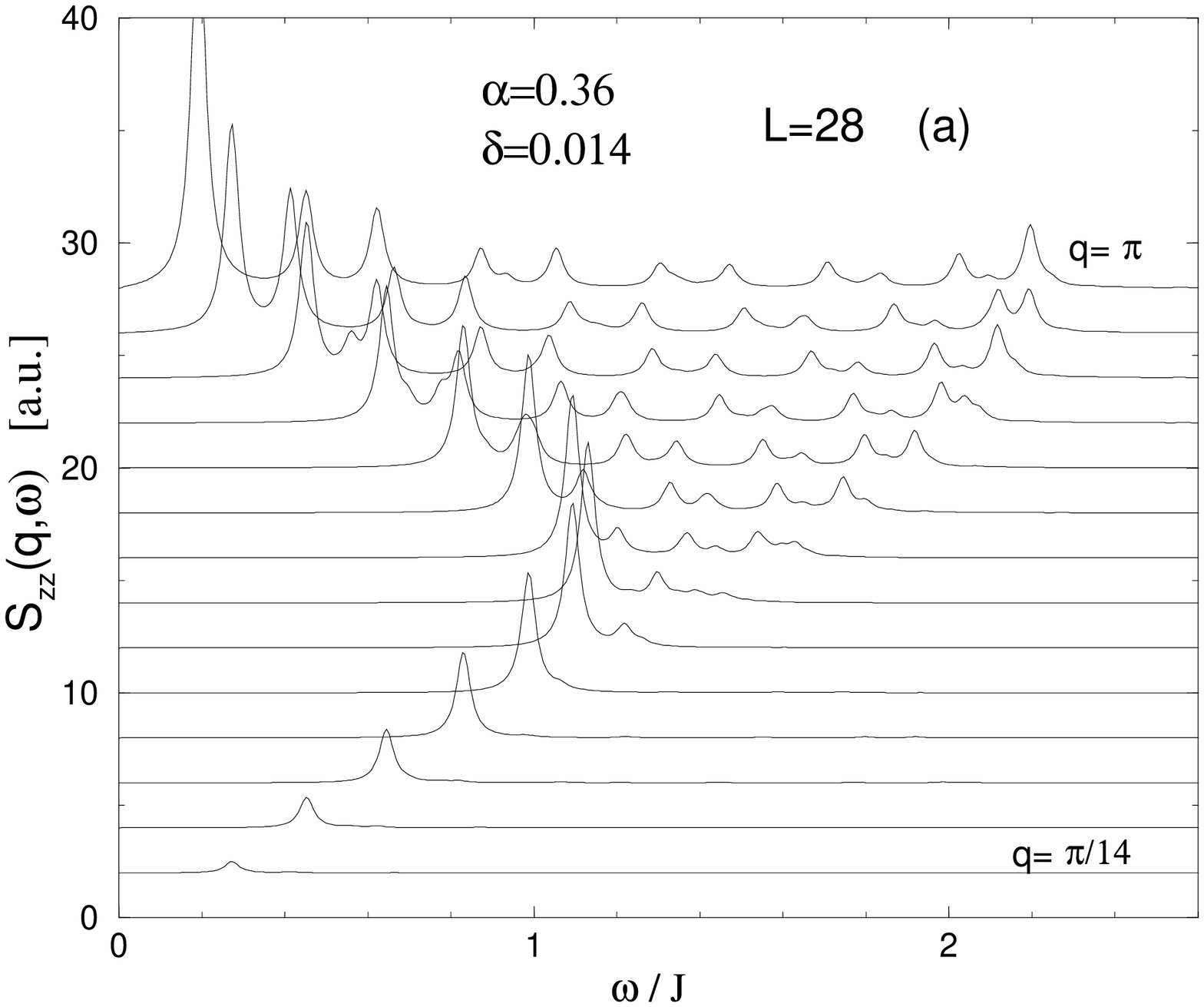,width=7cm,height=6.5cm,angle=-90}}
\mbox{\psfig{figure=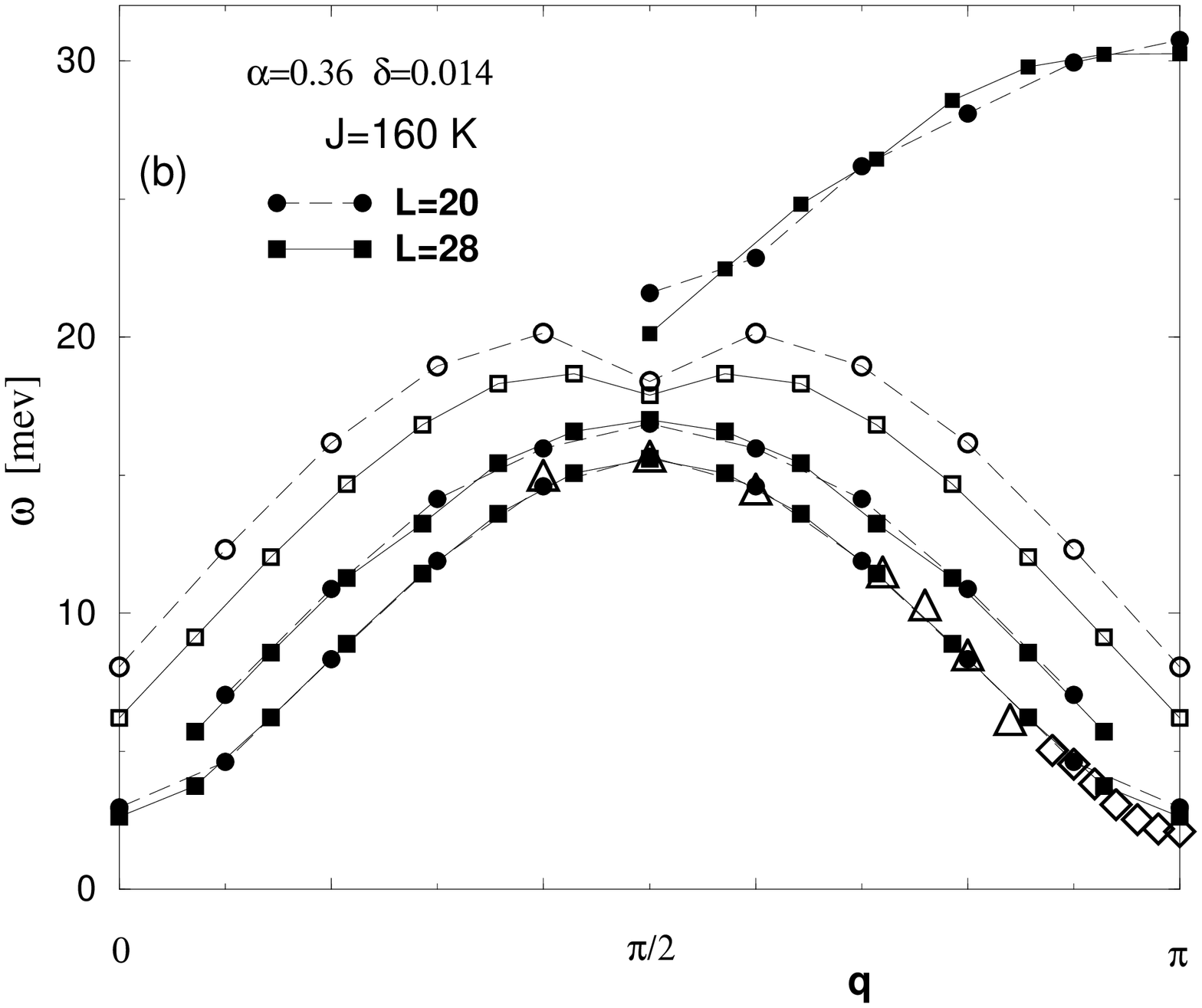,width=7cm,height=6.5cm,angle=-90}}
\end{center}
\caption{(a) Dynamical spectral function $S_{zz}(q,\omega)$ calculated at
momenta $q=n\frac{2\pi}{L}$ ($L=28$) in the D-phase. From bottom to top,
n goes from 1 to $L/2$ ($=14$) and the parameters are indicated 
on the plot. (b) Momentum--dependence of the various structures appearing
in $S_{zz}(q,\omega)$ for the same parameters as in (a). 
Results for both $20$- and $28$-sites clusters are shown. 
The open triangles and diamonds are experimental data taken from
Ref. \protect\onlinecite{Regnault}. }
\label{f1}
\end{figure}

In the D phase the transverse and longitudinal 
dynamical spin structure factor are given by
\begin{eqnarray}
S_{+-} (q,\omega,H)&=&
\sum_n |\big<\Psi_n|S_q^+|\Psi_0\big>|^2
\delta(\omega-E_n+H-E_0)\ , \cr
S_{zz} (q,\omega,H) &\equiv& S_{zz} (q,\omega)
                           =\frac{1}{2}S_{+-} (q,\omega,0)\ ,
\end{eqnarray}
where $\Psi_0$ is the $S_z=0$ ground state and the sum over $n$ is, in fact, 
restricted to the $S_z=1$ components of the excited triplet states. 
Hereafter, the units of the magnetic field $H$ are such that $g\mu_B=1$. 
The longitudinal part is field independent while the field dependence of the
tranverse part is simply due to the Zeeman splitting of the S=1 multiplet
producing a shift of the poles of the $\delta$-function. 
\vskip -1.0cm
\begin{figure}[htb]
\begin{center}
\mbox{\psfig{figure=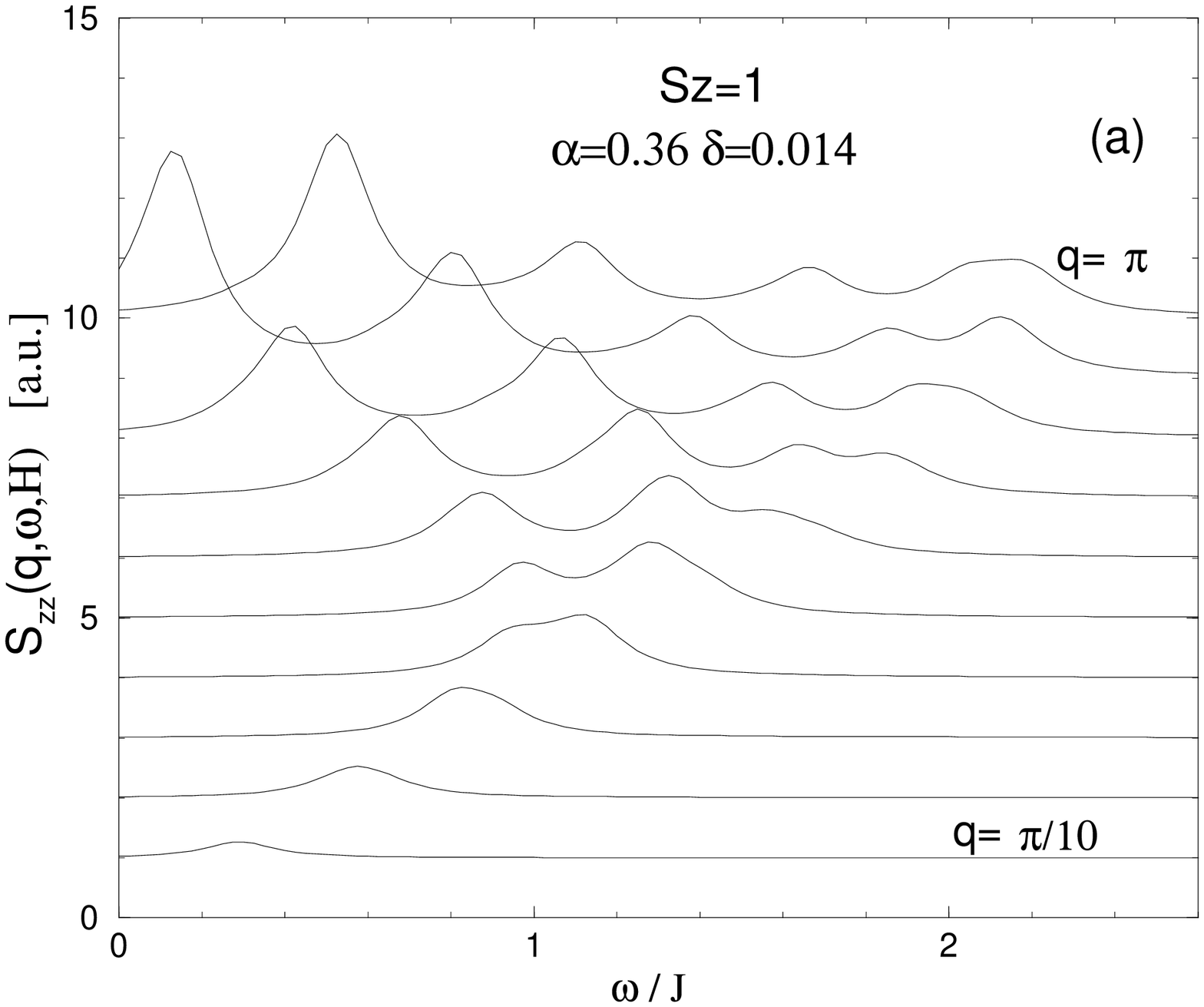,width=7cm,height=6.5cm,angle=-90}}
\mbox{\psfig{figure=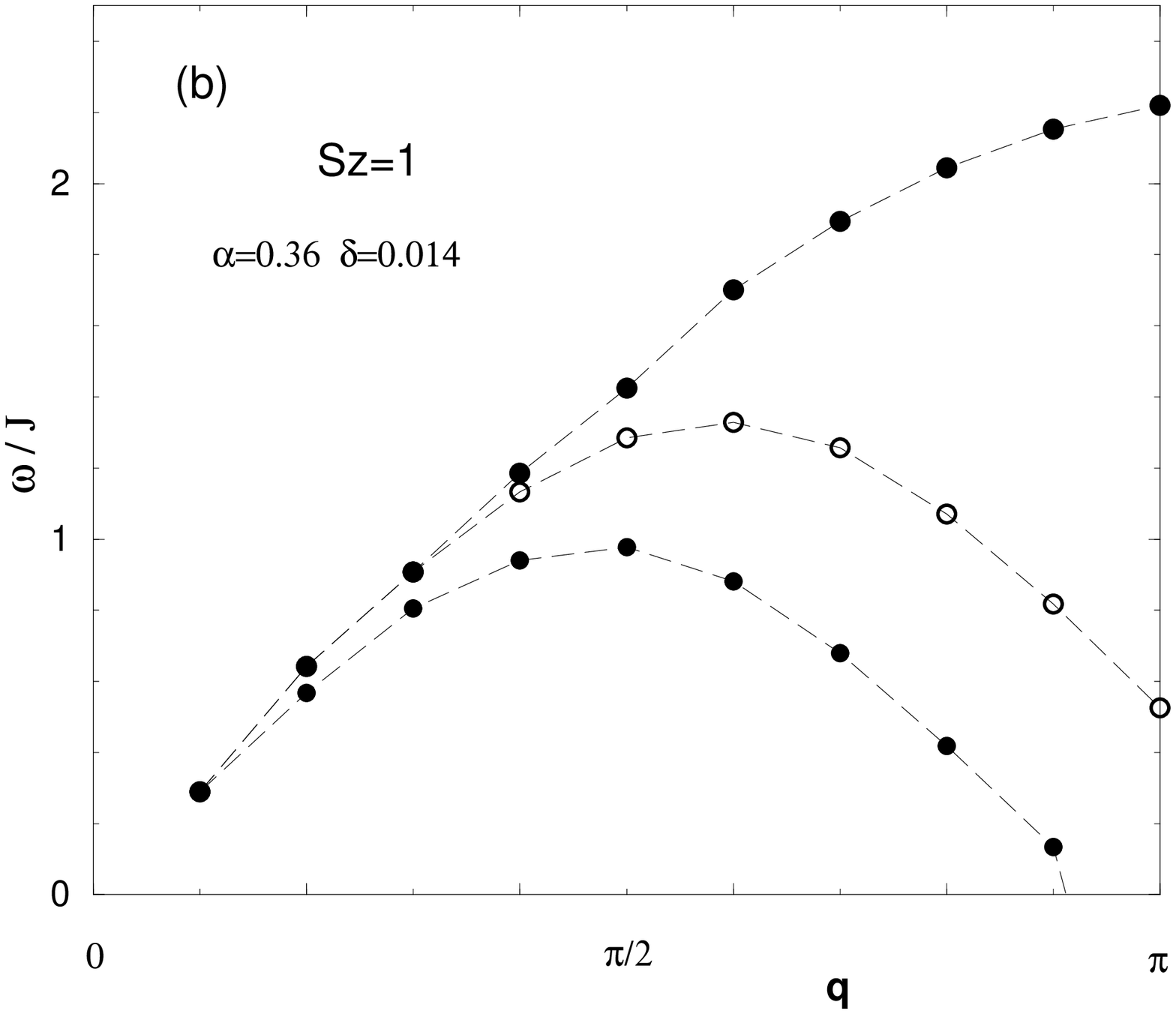,width=7cm,height=6.5cm,angle=-90}}
\end{center}
\caption{(a) Dynamical spectral functions $S_{zz}(q,\omega,H)$,
calculated at momenta \protect$q=2n\frac{\pi}{L}$ ($L=20$) in the 
$S_z=1$ ground state
(for $H>H_c$) (b) Momentum--dependence of the two main structures with lower
energy in $S_{zz}(q,\omega)$ and upper bound of the continuum
for the same parameters as in (a). }
\label{f2}
\end{figure}

Before discussing in more details the role of the magnetic field, we
briefly summarize our main results on the field independent longitudinal
structure factor. The results for $S_{zz}(q,\omega)$ 
calculated on a 28-sites chain 
are shown in Fig. \ref{f1}(a). A sharp low energy feature is seen which 
has been interpreted as a magnon branch or, equivalently as a
spinon-spinon bound state. A lot of weight is also seen
above this band. It is interesting to notice that a second sharp
peak exists above the magnon branch and might {\it a priori} be interpreted 
as a second bound state below the continuum \cite{Schulz}. The excitation 
energy of the three lowest triplet states as a function of momentum q is 
shown, for two different chain lengths, in Fig. \ref{f1}(b) together with the 
upper limit of the continuum of excitations. It is clear that, in contrast to 
the two first triplet states, the excitation energy of the third
one (open symbols) is strongly size dependent and, very likely, converges
to the value of the second excitation energy. This means that
the second and third excited states belong to a continuum of states. 
We conclude that, for these parameters, there is only
a single magnon branch separated from the continuum by 
a gap as seen experimentally \cite{Ain}. For completeness, we also show a 
comparison of the magnon dispersion with the available experimental data 
of Regnault et al. \cite{Regnault}. The agreement is excellent, even slightly
better than the fit obtained by Haas et al. \cite{Haas} for the
parameters proposed in Ref. \onlinecite{Castilla}. Note that, although
finite size effects are negligeable around $q=\pi/2$, a careful 
finite size scaling of the gap at $q=\pi$ (see Ref.~\onlinecite{Riera_field})
is necessary and, in fact, gives a gap very close to the experimental value. 
\vskip -1.0cm 
\begin{figure}[htb]
\begin{center}
\mbox{\psfig{figure=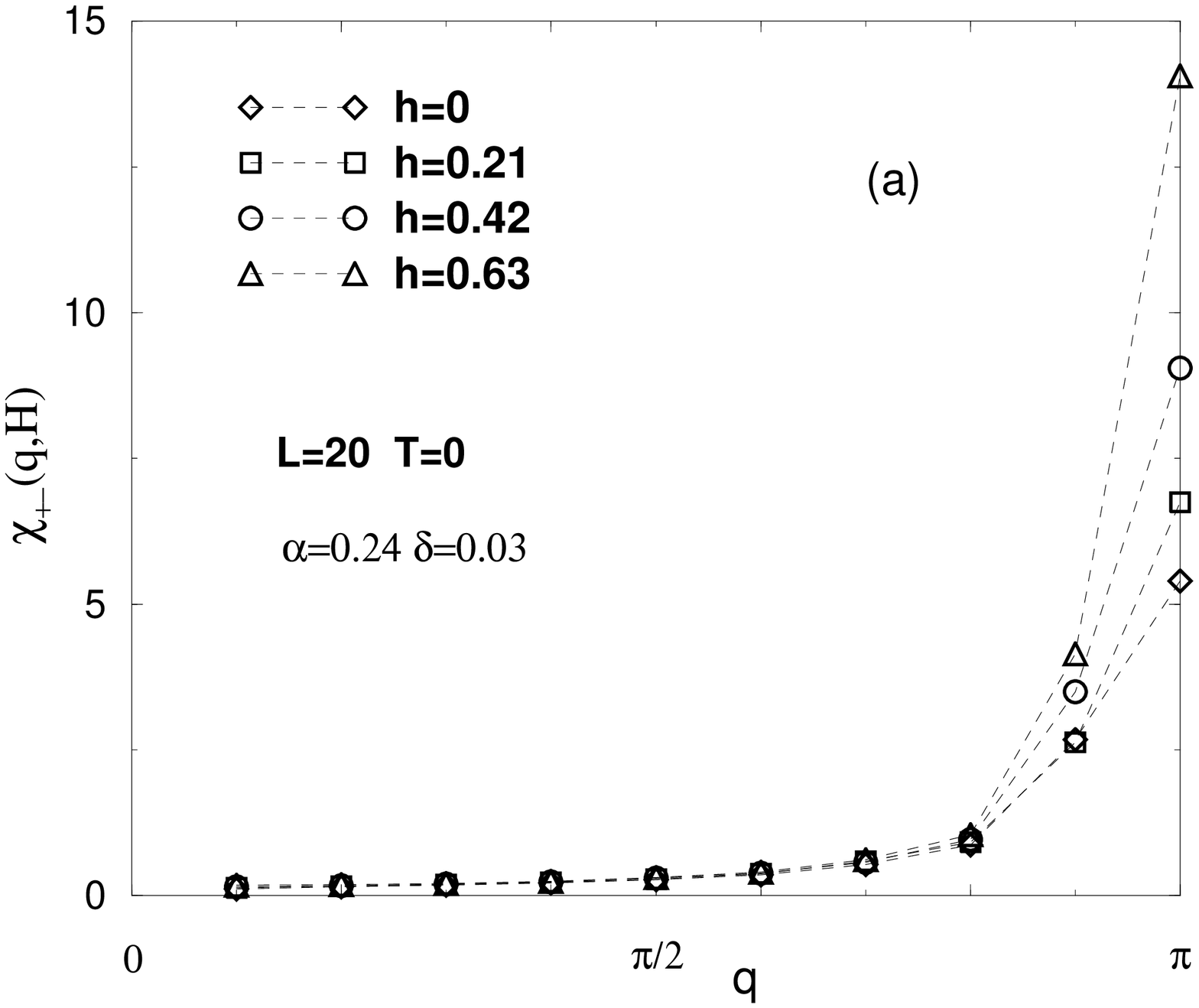,width=7cm,height=6.5cm,angle=-90}}
\mbox{\psfig{figure=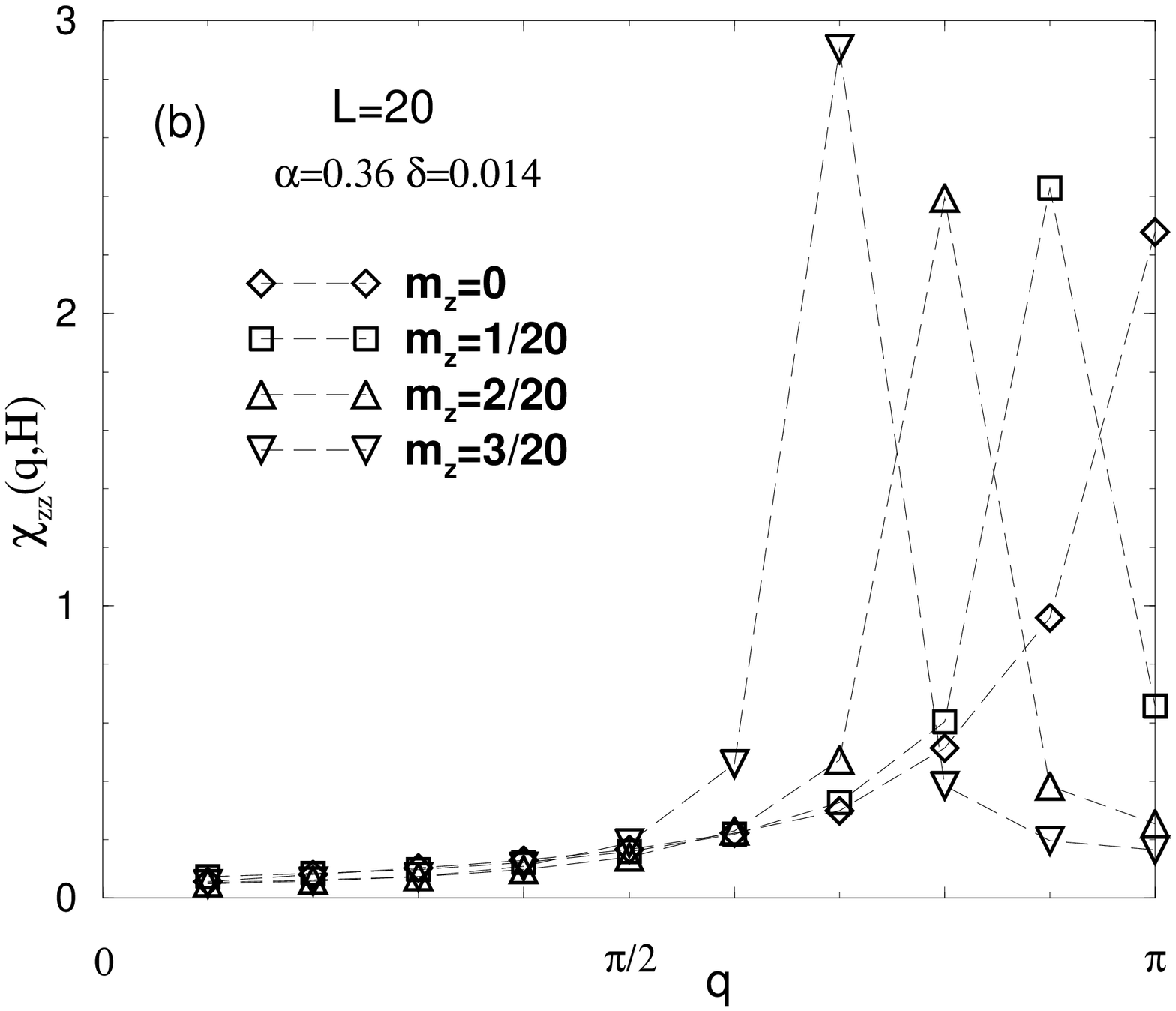,width=7cm,height=6.5cm,angle=-90}}
\end{center}
\caption{Momentum dependence of static susceptibilities. (a) $\chi_{+-}$ 
for several values of the reduced fields $h=H/H_c$. (b) $\chi_{zz}$ 
for several values of the magnetisation $m_z$. 
Parameters are indicated on the graph.}
\label{f3}
\end{figure}

We now turn to the discussion of the effect of the magnetic field H.
For $H<H_c$ ($H_c\sim \Delta_S$), the transverse part $S_{+-}$
is simply obtained by a 
global shift of $H$ of the spectral weight of Fig. \ref{f1}(a) towards low 
energies. At $H_c$ the GS acquires a finite magnetisation $S_z$ 
and the lattice modulation becomes incommenturate. In
the following, we shall use a sinusoidal modulation 
$J(i)=J(1+\delta\cos{(K_{\delta}\, i)})$ of momentum 
$K_{\delta}=\frac{2\pi}{L}(L/2-S_z)$ which minimizes the total energy.
However, we stress that the effects directly connected 
to the change of the lattice 
modulation are very weak \cite{note_soliton}.
In the GS carrying a finite magnetisation the longitudinal dynamical spin 
structure factor becomes
\begin{eqnarray}
S_{zz} (q,\omega,H)&=&
\sum_n |\big<\Psi_n|S_q^Z|\Psi_0(S_z)\big>|^2
\delta(\omega-E_n-E_0)\ ,
\end{eqnarray}
where the field dependence enters in the GS $|\Psi_0(S_z)\big>$.
Results are shown in Fig. \ref{f2}(a-b) for $S_z=1$. The low energy spectral 
weight appears now away from momentum $\pi$ at a momentum $q_{\text{min}}(H)$. 
Its position (see Fig. \ref{f2}(b))
in q-space shifts continuously with the field $H$ according to
$q_{\text{min}}(H)=\pi(1-2m_z)$ where $m_z=S_z/L$ is the magnetization.
This is consistent with the mapping of the XY model on the non-interacting 
tight-binding fermion model, the change of the magnetic field being
related, in the fermion picture, to a change of the chemical potential
and hence of the Fermi momentum $k_F$ (Ref. \onlinecite{theory_old}). 

Besides inelastic neutron scattering, Nuclear Magnetic Resonance (NMR) is a
very efficient technique to probe the spin dynamics~\cite{Giam}, 
and in particular, the
static (zero frequency) susceptibility tensor $\chi _{\alpha}(q)$.
The dominant component of the
interaction between nuclear spins of copper in CuGeO$_3$ is namely the one
mediated by the fluctuation of electronic spins, i.e., by the non-local
electron spin susceptibility. Neglecting the direct nuclear dipole - dipole
interaction, the second moment ($M_2$) of the NMR lineshape obtained in the
standard spin - echo experiment\cite{M2=T2G} will in general\cite{T2formula}
depend on both $\chi _{zz}$ and $\chi _{+-}$ susceptibilities. However,
taking advantage of the strong anisotropy of the copper nuclear spin
hyperfine coupling $(A_{ZZ}/A_{\perp })^2$ $\simeq $ 100,
(where $Z$ is the principal axis perpendicular to the
$d_{x^2-y^2}$ orbital of copper spin),\cite{NMR_PRB} we can select only
one dominant contribution to $M_2$ by applying the magnetic field in the
appropriate direction. For $H\parallel Z$ or $H\parallel $ to the chain axis,
we thus measure $M_{2,zz}$  or $M_{2,+-}$, respectively, where

\begin{eqnarray}
M_{2,\alpha}=C_\alpha \{\frac 1L\sum_qA_{ZZ}^4\chi _\alpha ^2(q)
-[\frac 1L\sum_qA_{ZZ}^2\chi _\alpha (q)]^2\}\ ,
\label{second_moment}
\end{eqnarray}
with $\alpha =zz$ or $+-$, $C_{zz}=0.69(\hbar \gamma _n/g\mu _B)^4/8\hbar ^2$,
$C_{+-}=pC_{zz}$ for the (-1/2, 1/2) transition and $C_{+-}=9pC_{zz}/16$ for
$(\pm$3/2, $\pm$1/2) transitions, where $p\leq 1$ is the proportion of
''like'' nuclear spins participating in spin-exchange interaction
\cite{T2formula}. The $q$-dependence, $A_{ZZ}=A_{ZZ}(q)$,
of the $A_{ZZ}$ coupling was found to be essentially negligible
($\leq $ 10\%) with $A_{ZZ}(q=\pi )\simeq -440$ kOe.\cite{NMR_PRB}

The static susceptibilities can be easily obtained from the knowledge of the
dynamical structure factors,
\begin{eqnarray}
\chi_{\alpha}(q,H)=\int_{0}^\infty \frac{S_{\alpha}(q,\omega,H)}{\omega}\, d\omega \ .
\end{eqnarray}
Results obtained on a 20-sites cluster are shown in Fig. \ref{f3}(a-b).
In the D-phase ($h<1$, $m_z=0$) $\chi_{zz}(q)$ and  $\chi_{+-}(q)$ are peaked 
at momentum $q=\pi$. For increasing field, $\chi_{zz}(q)$ remains constant
($\chi_{zz}(q,H)=\frac{1}{2}\chi_{+-}(q,H=0)$) while a strong singularity 
developps at $q=\pi$ in $\chi_{+-}(q,H)$ when $h\rightarrow 1$. 
Above the critical point ($h>1$), the longitudinal part $\chi_{zz}(q,H)$
becomes field dependent with a maximum at the momentum $q_{\text{min}}$
corresponding to the appearance of spectral weight at low energies
in the dynamical spin structure factor.
\vskip -1.0cm 
\begin{figure}[htb]
\begin{center}
\mbox{\psfig{figure=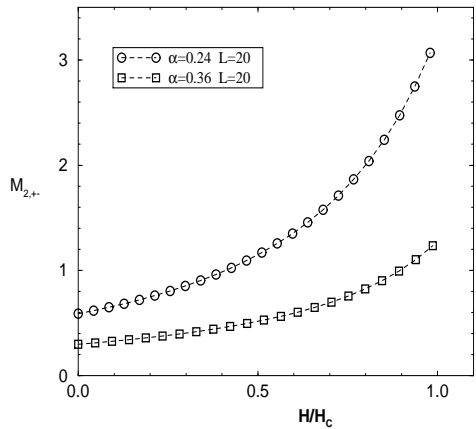,width=7cm,height=6.5cm,angle=-90}}
\end{center}
\caption{Second moment $M_{2,+-}$ vs reduced field $H/H_c$.
To avoid spurious divergences
when $h\rightarrow 1$, the momentum $q=\pi$ is not included in the
corresponding sum of Eq.~\protect\ref{second_moment}.}
\label{f4}
\end{figure}

The field dependence of the susceptibilities are expected to be directly
reflected in the copper NMR data in CuGeO$_3$. The $M_{2,zz}$ value is thus
predicted to be field independent, while $M_{2,+-}$ should strongly grow
as we increase $h\rightarrow 1$. (The increase of $M_{2,+-}(h)$ is plotted
in Fig. \ref{f4} in reduced units, i.e., with $C_\alpha =1=A_{ZZ}(q)$,
as the value of $p$ is not known). Indeed, NMR spectra in the D-phase,
i.e., the width of NMR lines (at 4.2 K) reported 
in Ref.~\protect\onlinecite{NMR_PRB},
seem to confirm this behavior. These results have to be
corroborated by further systematic investigations.

In conclusion, we have shown that the strong magnetic field dependence of
the spin fluctuations could lead to new experimentally observable effects.
In particular, it is suggested that the behavior of the transverse spin
susceptibility can be experimentally extracted from the second moment of the
NMR spectrum.

We thank IDRIS (Orsay) 
for allocation of CPU time on the C94 and C98 CRAY supercomputers.


\begin{references}
\bibitem{comment_interchain} 
The weak interchain coupling ($J_\perp/J < 0.1$) is neglected here.
\end{references}
\end{document}